\documentclass[showpacs,preprintnumbers,amsmath,amssymb,prb]{revtex4}

\usepackage{graphicx}
\usepackage{dcolumn}
\usepackage{bm}

\begin{document}

\title{Renormalization of the spin-wave spectrum in three-dimensional ferromagnets with dipolar interaction}

\author{A. V. Syromyatnikov}
 \email{syromyat@thd.pnpi.spb.ru}
\affiliation{Petersburg Nuclear Physics Institute, Gatchina, St.\ Petersburg 188300, Russia}

\date{\today}

\begin{abstract}

Renormalization of the spin-wave spectrum is discussed in a cubic ferromagnet with dipolar forces at $T_C\gg T\ge0$. First $1/S$-corrections are considered in detail to the bare spectrum $\epsilon_{\bf k} = \sqrt{Dk^2 (Dk^2 + S\omega_0\sin^2\theta_{\bf k})}$, where $D$ is the spin-wave stiffness, $\theta_{\bf k}$ is the angle between $\bf k$ and the magnetization and $\omega_0$ is the characteristic dipolar energy. In accordance with previous results we obtain the thermal renormalization of constants $D$ and $\omega_0$ in the expression for the bare spectrum. Besides, a number of previously unknown features are revealed. We observe terms which depend on azimuthal angle of the momentum $\bf k$. They are proportional to the square invariant $|k_x^2 - k_y^2|/k$ in the case of simple cubic lattice and to the triangular invariant ${\rm Im}([k_x + ik_y]^3)/k^2$ in face- and body-centered cubic lattices, where we assume that $z$-axis is directed along magnetization. It is obtained an isotropic term proportional to $k$ which makes the spectrum linear rather than quadratic when $\sin\theta_{\bf k}=0$ and $k \ll \omega_0/D$. In particular a spin-wave gap proportional to $\sin\theta_{\bf k}$ is observed. Essentially, thermal contribution from the Hartree-Fock diagram to the isotropic correction as well as to the spin-wave gap are proportional to the demagnetizing factor in the direction of domain magnetization. This nontrivial behavior is attributed to the long-range nature of the dipolar interaction. It is shown that the gap screens infrared singularities of the first $1/S$-corrections to the spin-wave stiffness and longitudinal dynamical spin susceptibility (LDSS) obtained before. We demonstrate that higher order $1/S$-corrections to these quantities are small at $T\ll\omega_0$. However the analysis of the entire perturbation series is still required to derive the spectrum and LDSS when $T\gg\omega_0$. Such an analysis is out of the scope of the present paper. 

\end{abstract}


\maketitle

\section{Introduction}
\label{int}

Despite enormous amount of studies on the theory of ferromagnets with magnetic dipolar interaction there are still some unresolved problems. Infrared divergences of the longitudinal dynamical spin susceptibility \cite{top,luz} and infrared divergences of $1/S$-corrections to the spin-wave stiffness \cite{rahman} are among them.

It was obtained by Toperverg and Yashenkin \cite{top} that dipolar forces lead to strong long-wavelength fluctuations manifesting themselves in infrared divergence of the first perturbation corrections to the uniform longitudinal spin susceptibility: $\chi_\|(\omega\to0) \sim iT/\omega$. It is pointed out in Ref.~\cite{top} that this singularity is nonphysical one since it leads to a nonzero value of absorption function $Q_\omega\propto\omega{\rm Im}\chi(\omega)$ at $\omega=0$ and thus the sample would be heated by a dc field. Then one could expect great renormalization of $\chi_\|(\omega,{\bf k})$ after taking into account higher order perturbation corrections at small energy and momentum. Thus, experiment on nearly isotropic ferromagnet CdGr$_2$Se$_4$ shown weaker power-law dependence: \cite{luz} $\chi_\|(\omega\to0) \sim (i/\omega)^{0.28}$. The analysis of the entire perturbation series for $\chi_\|(\omega,{\bf k})$ has not been carried out yet.

First $1/S$-corrections to the spin-wave spectrum have been discussed by Rahman and Mills in Ref.~\cite{rahman}. They found an infrared divergent contribution to the spin-wave stiffness. It stems from three-magnons terms in the Hamiltonian. As they did not obtain any term remaining finite as $k\to0$, one could expect a great renormalization of the spectrum at small momenta. However Rahman and Mills used an approximation discussed below in detail to simplify intermediate calculations. It remains unclear if there exists a constant term as $k\to0$ outside this approximation.

These problems of infrared singularities are related to the fact that a spin-wave gap has not been obtained yet in magnon spectrum. It is well known that within the linear spin-wave approximation the spectrum has the following form at small $k$: \cite{sw,kittel}
\begin{equation}
\label{spec0}
\epsilon_{\bf k} = \sqrt{\left(Dk^2 + g\mu H^{(i)}\right)\left(Dk^2 + g\mu H^{(i)} + S\omega_0\sin^2\theta_{\bf k}\right)},
\end{equation}
where $D$ is the spin-wave stiffness, $H^{(i)}$ is the intrinsic magnetic field, $\theta_{\bf k}$ is the angle between $\bf k$ and the magnetization and
\begin{equation}
\label{o0}
\omega_0 = 4\pi \frac{(g\mu)^2}{v_0}
\end{equation}
is the characteristic dipolar energy, where $v_0$ is the unit cell volume. It is well established that a finite-size sample of a ferromagnet breaks up into domains at zero external magnetic field so as the intrinsic magnetic field is zero. \cite{landau} It is seen from Eq.~(\ref{spec0}) that the spectrum has no gap in this case. The intrinsic magnetic field is also zero in a unidomain sample that is infinite along the direction of magnetization.

At the same time there are some considerations in favor of existence of the spin-wave gap in a ferromagnet with dipolar forces. First, due to its long-range nature and symmetry the dipolar interaction violates the well-known Goldstone theorem and can lead to the gap. Second, it was obtained in Ref.~\cite{pet} that pseudodipolar interaction which has the same symmetry as the long-range dipolar interaction leads to an energy gap in antiferromagnets. Third, it is well-known that within the first order of $1/S$ dipolar and pseudodipolar forces lead to anisotropic corrections to the total energy of a ferromagnet. \cite{tes,vleck,kef_an} As a result, directions along edges of the cube are energetically favorable in a simple cubic lattice (SCL) whereas the magnetization should be parallel to a body diagonal of the cube in the face-centered cubic lattice (FCCL) and the body-centered cubic lattice (BCCL). \cite{keffer,tes,vleck,kef_an} It has been pointed out by Keffer \cite{keffer,kef_an} that the anisotropic terms in the total energy of a ferromagnet should be accompanied with an "energy shift" in the spin-wave spectrum. However the particular calculations have not been realized yet thoroughly because of their cumbersomeness. 

In the present paper we carry out a comprehensive analysis of the real part of first $1/S$-corrections to the spectrum of a ferromagnet with a cubic lattice. We use a technique based on introduction of anomalous Green's functions rather than on conventional Bogolyubov's transformation. This technique proved to be more convenient making intermediate calculations more compact. \cite{pet,malold,syromyat,chir} In accordance with Ref.~\cite{rahman} we obtain the thermal renormalization of constants $D$ and $\omega_0$ in Eq.~(\ref{spec0}). Besides, we reveal a number of previously unknown features. In particular it is shown that there is a gap in the spectrum proportional to $\sin\theta_{\bf k}$. This gap screens singularities of $1/S$-corrections to the spin-wave stiffness and longitudinal spin susceptibility. We show that higher order $1/S$-corrections to these quantities are small at $T\ll\omega_0$. However the analysis of the entire perturbation series is still required when $T\gg\omega_0$ which is out of the scope of the present paper. In this case one could expect large renormalization of the longitudinal dynamical susceptibility as the nontrivial power-law dependence of $\chi_\|(\omega)$ discussed above was obtained experimentally in CdCr$_2$Se$_4$ at $T\gg\omega_0$. \cite{luz}

Then, we observe corrections to the spectrum which depend on azimuthal angle of the momentum $\bf k$. They are proportional to the square invariant $|k_x^2 - k_y^2|/k$ in the case of SCL and to the triangular invariant ${\rm Im}([k_x + ik_y]^3)/k^2$ for BCCL and FCCL, where we assume that $z$-axis is directed along magnetization. This difference is accounted for the fact that due to the dipolar anisotropy magnetization is parallel to an edge of a cube (tetrad axis) in SCL and to a body diagonal of a cube (triad axis) in BCCL and FCCL. It is also shown that there is an isotropic term proportional to $k$ that plays the predominant part when $\sin\theta_{\bf k}=0$ and $k\ll\omega_0/D$: renormalized spectrum is linear in $k$ rather than quadratic [cf.\ Eq.~(\ref{spec0})]. We demonstrate a relation between the dipolar anisotropy and the gap and isotropic term in the spectrum. Both the gap and the isotropic term are proportional to sums over momenta containing anomalous Green's functions that were omitted in Ref.~\cite{rahman}.

Remarkably, thermal contribution from the Hartree-Fock diagram to the isotropic correction as well as to the spin-wave gap are proportional to the demagnetizing factor in the direction of the domain magnetization. Then a spin-wave excitation with wavelength much smaller than the size of the domain appears to "feel" the domain shape. This nontrivial behavior is attributed to the long-range nature of the dipolar interaction.

The rest of this paper is organized as follows. The Hamiltonian transformation and the technique are discussed in Sec.~\ref{hamtr}. First $1/S$-corrections to the spin-wave spectrum are considered in Sec.~\ref{specorr}. In Sec.~\ref{disc} we discuss the screening of infrared singularities of longitudinal spin fluctuations and further $1/S$-corrections to the spin-wave stiffness. Possibility of experimental verification of the results obtained is also considered in Sec.~\ref{disc}. Sec.~\ref{con} contains our conclusions. One appendix is included with details of calculations.

\section{Hamiltonian transformation}
\label{hamtr}

The Hamiltonian of a ferromagnet with dipolar interaction in magnetic field $\bf H$ has the well known form:
\begin{eqnarray}
\label{ham0}
{\cal H} &=& -\frac12 \sum_{l\ne m} (J_{lm}\delta_{\alpha\beta} + Q_{lm}^{\alpha\beta}) S_l^\alpha  S_m^\beta +g\mu H \sum_l S_l^z,\\
\label{q}
Q_{lm}^{\alpha\beta} &=& (g\mu)^2\frac{3R_{lm}^\alpha R_{lm}^\beta - \delta_{\alpha\beta}R_{lm}^2}{R_{lm}^5}.
\end{eqnarray}
We assume that $H=0$ in the case of the unidomain sample that is infinite in the direction of magnetization. For multidomain sample the Hamiltonian (\ref{ham0}) describes one domain and $H$ is the magnetic field produced by all other domains. Taking the Fourier transformation we have from Eq.~(\ref{ham0}):
\begin{equation}
\label{ham}
{\cal H} = -\frac12 \sum_{\bf k}(J_{\bf k}\delta_{\alpha\beta} + Q_{\bf k}^{\alpha\beta}) S_{\bf k}^\alpha  S_{-\bf k}^\beta +g\mu H  \mathfrak N S_{\bf 0}^z,
\end{equation}
where $J_{\bf k} = \sum_l J_{lm}\exp(i{\bf k R}_{lm})$, $Q_{\bf k}^{\alpha\beta} = \sum_l Q_{lm}^{\alpha\beta}\exp(i{\bf k R}_{lm})$ and $\mathfrak N$ is the number of spins in the lattice. Dipolar tensor $Q_{\bf k}^{\alpha\beta}$ possesses the well known properties which are the same for all types of cubic lattices and independent of the orientation of the coordinate frame relative to cubic axes: \cite{sums,sw} 
\begin{eqnarray}
\label{qsmall}
Q_{\bf k}^{\alpha\beta} &=& \omega_0\left( \frac {\delta_{\alpha\beta}}{3} - \frac{k_\alpha k_\beta}{k^2} \right),  
\mbox{ if } \frac 1L\ll k\ll1,\\
\label{q0}
Q_{\bf 0}^{\alpha\beta} &=& \omega_0 \left(\frac13 - {\cal N}_\alpha\right) \delta_{\alpha\beta},
\end{eqnarray}
where we set the lattice spacing equal to unity, $L$ is the characteristic length of the domain, ${\cal N}_\alpha$ are demagnetizing factors and $\omega_0$ is the characteristic dipolar energy given by Eq.~(\ref{o0}). 

After Dyson-Maleev transformation
\begin{equation}
\label{md}
S^x_{\bf k} = \sqrt{\frac S2} \left( a_{\bf k} + a^\dagger_{-\bf k} - \frac{(a^\dagger a^2)_{\bf k}}{2S} \right), \qquad 
S^y_{\bf k} = -i\sqrt{\frac S2} \left( a_{\bf k} - a^\dagger_{-\bf k} - \frac{(a^\dagger a^2)_{\bf k}}{2S} \right), \qquad  
S^z_{\bf k} = S - (a^\dagger a)_{\bf k}
\end{equation}
Hamiltonian (\ref{ham}) has the form ${\cal H} = E_0 + \sum_{i=1}^6 {\cal H}_i$, where $E_0$ is the ground state energy and ${\cal H}_i$ denote terms containing products of $i$ operators $a$ and $a^\dagger$. As we intend to calculate corrections of the first order of $1/S$, we need terms up to ${\cal H}_4$. ${\cal H}_1=0$ because it contains only $Q_{\bf 0}^{\alpha\beta}$ with $\alpha\ne\beta$. For the rest necessary terms one has:
\begin{eqnarray}
\label{h2}
{\cal H}_2 &=& \sum_{\bf k} \left[E_{\bf k} a^\dagger_{\bf k}a_{\bf k} + \frac{B_{\bf k}}{2} a_{\bf k}a_{-\bf k} + 
\frac{B_{\bf k}^*}{2} a^\dagger_{\bf k}a^\dagger_{-\bf k}\right],\\
\label{h3}
{\cal H}_3 &=& \sqrt{\frac {S}{2 \mathfrak N}} \sum_{{\bf k}_1 + {\bf k}_2 + {\bf k}_3 = 0} a^\dagger_{-1}\left[a^\dagger_{-2}(Q_2^{xz}+iQ_2^{yz}) + a_2(Q_2^{xz} - iQ_2^{yz})\right]a_3,\\
{\cal H}_4 &=& \frac{1}{4 \mathfrak N}\sum_{{\bf k}_1 + {\bf k}_2 + {\bf k}_3 + {\bf k}_4 = 0} 
\left\{
2(J_1-J_{1+3})a^\dagger_{-1}a^\dagger_{-2}a_3a_4 \right.\nonumber\\
\label{h4}
&&{}\left.
+ a^\dagger_{-1} \left[a_2 (Q_2^{xx} - 2iQ_2^{xy} - Q_2^{yy}) + a^\dagger_{-2}( Q_2^{xx} + Q_2^{yy} - 2Q_{2+3}^{zz}) \right] a_3a_4 \right\},
\end{eqnarray}
where we drop index $\bf k$ in Eqs.~(\ref{h3}) and (\ref{h4}) and
\begin{eqnarray}
\label{e}
E_{\bf k} &=& S(J_{\bf 0} - J_{\bf k}) - \frac S2\left( Q_{\bf k}^{xx} + Q_{\bf k}^{yy} - \frac{2\omega_0}{3} \right) -g\mu(4\pi g\mu S{\cal N}_z + H) \nonumber\\
& \stackrel{k\ll1}{\approx} & Dk^2 + \frac{S\omega_0}{2}\sin^2\theta_{\bf k} - g\mu(H + 4\pi g\mu S{\cal N}_z),
\\
\label{b}
B_{\bf k} &=& -\frac S2 \left( Q_{\bf k}^{xx} - 2iQ_{\bf k}^{xy} - Q_{\bf k}^{yy} \right)
\stackrel{k\ll1}{\approx} \frac{S\omega_0}{2}\sin^2\theta_{\bf k}e^{-2i\phi_{\bf k}},
\end{eqnarray}
where the expressions after $\stackrel{k\ll1}{\approx}$ are approximate values of corresponding quantities at $k\ll1$. In the multidomain sample the expression in the brackets of the last term in Eq.~(\ref{e}) is the intrinsic magnetic field which is zero. In the unidomain sample that is infinite in the direction of magnetization we have ${\cal N}_z=0$, $H=0$ and the last term in Eq.~(\ref{e}) also vanishes. In the spin-wave approximation the magnon spectrum is 
\begin{equation}
\label{spec1}
\epsilon_{\bf k} = \sqrt{E_{\bf k}^2 - |B_{\bf k}|^2}. 
\end{equation}
Using Eqs.~(\ref{qsmall}), (\ref{e}) and (\ref{b}) we obtain from Eq.~(\ref{spec1}) at $L^{-1}\ll k\ll1$ the well known result (\ref{spec0}).

To perform the calculations it is convenient to introduce the following retarded Green's functions: $G(\omega,{\bf k}) = \langle a_{\bf k}, a^\dagger_{\bf k} \rangle_\omega$, $F(\omega,{\bf k}) = \langle a_{\bf k}, a_{-\bf k} \rangle_\omega$, ${\overline G}(\omega,{\bf k}) = \langle a^\dagger_{-\bf k}, a_{-\bf k} \rangle_\omega = G^*(-\omega,-{\bf k})$ and $F^\dagger (\omega,{\bf k}) = \langle a^\dagger_{-\bf k}, a^\dagger_{\bf k} \rangle_\omega = F^*(-\omega,-{\bf k})$. We have two sets of Dyson equations for them. One of these sets has the form:
\begin{equation}
\label{eqfunc}
\begin{array}{l}
G(\omega,{\bf k}) = G^{(0)}(\omega,{\bf k}) + G^{(0)}(\omega,{\bf k}){\overline \Sigma}(\omega,{\bf k})G(\omega,{\bf k}) + G^{(0)}(\omega,{\bf k}) [B_{\bf k}^* + \Pi(\omega,{\bf k})] F^\dagger(\omega,{\bf k}),\\
F^\dagger(\omega,{\bf k}) = {\overline G}^{(0)}(\omega,{\bf k}) \Sigma(\omega,{\bf k})F^\dagger(\omega,{\bf k}) + {\overline G}^{(0)}(\omega,{\bf k}) [B_{\bf k} + \Pi^\dagger(\omega,{\bf k}) ]G(\omega,{\bf k}),
\end{array}
\end{equation}
where $G^{(0)}(\omega,{\bf k}) = (\omega - E_{\bf k}+i\delta)^{-1}$ is the bare Green's function and $\Sigma$, $\overline \Sigma$, $\Pi$ and $\Pi^\dagger$ are the self-energy parts. Solving Eqs.~(\ref{eqfunc}) one obtains:
\begin{eqnarray}
G(\omega,{\bf k}) &=& \frac{\omega + E_{\bf k} + \Sigma(\omega,{\bf k})}{{\cal D}(\omega,{\bf k})},\nonumber\\
F(\omega,{\bf k}) &=& -\frac{B^*_{\bf k} + \Pi(\omega,{\bf k})}{{\cal D}(\omega,{\bf k})},\nonumber\\
\label{gf}
{\overline G}(\omega,{\bf k}) &=& \frac{-\omega + E_{\bf k} + {\overline \Sigma}(\omega,{\bf k})}{{\cal D}(\omega,{\bf k})},\\
F^\dagger(\omega,{\bf k}) &=& -\frac{B_{\bf k} + \Pi^\dagger(\omega,{\bf k})}{{\cal D}(\omega,{\bf k})},\nonumber
\end{eqnarray}
where
\begin{equation}
\label{d}
{\cal D}(\omega,{\bf k}) = (\omega+i\delta)^2 - \epsilon_{\bf k}^2 - E_{\bf k}(\Sigma + \overline{\Sigma}) + B_{\bf k}\Pi + B^*_{\bf k}\Pi^\dagger + (\omega + i\delta)(\Sigma - \overline{\Sigma}) + \Pi\Pi^\dagger - \Sigma \overline{\Sigma}
\end{equation}
and $\epsilon_{\bf k}$ is given by Eq.~(\ref{spec1}). The last two terms in Eq.~(\ref{d}) give corrections of at least second order of $1/S$ and will not be considered below. It is convenient for the following to introduce a new quantity $\Omega(\omega,{\bf k})$ describing the spin-wave spectrum renormalization [cf. Eq.~(\ref{d})]:
\begin{equation}
\label{o}
{\cal D}(\omega,{\bf k}) = (\omega+i\delta)^2 - \epsilon_{\bf k}^2 - \Omega(\omega,{\bf k}).
\end{equation}
We calculate $\Omega(\omega,{\bf k})$ in the next section.

\section{Renormalization of the spin-wave spectrum}
\label{specorr}

Corrections to the spin-wave spectrum to be obtained are proportional to sums over momenta containing components of the dipolar tensor $Q_{\bf k}^{\alpha\beta}$. In some of these sums summation over small momenta is important and one can use expressions (\ref{qsmall}) and (\ref{q0}) for $Q_{\bf k}^{\alpha\beta}$. As properties of the dipolar tensor (\ref{qsmall}) and (\ref{q0}) are the same for all types of cubic lattices and independent of orientation of the coordinate frame relative to the cubic axes, such sums do not depend on the direction of the quantized axis and the lattice type. Meantime there are sums in which summation over large momenta is essential and which, consequently, depend on the direction of quantized axis and the lattice type. Thus, one should bear in mind what is the direction of magnetization in the ground state. 

Therefore the well known fact should be taken into account that dipolar and pseudodipolar interactions lead the energy of a ferromagnet to be dependent on the direction of quantized axis. \cite{tes,kef_an} It has been shown in Ref.~\cite{tes} that the first $1/S$-correction to the energy $E_0$ gives rise to such anisotropy. In our notation this correction has the form:
\begin{equation}
\label{de}
\Delta E = \langle {\cal H}_2\rangle = \sum_{\bf k} \frac{\epsilon_{\bf k} - E_{\bf k}}{2} 
\approx - \sum_{\bf k} \frac{ |B_{\bf k}|^2}{4 \epsilon_{\bf k}},
\end{equation}
where the isotropic part of the energy is omitted in the right part of this expression. After direct calculations one obtains:
\begin{eqnarray}
\label{an}
\frac{\Delta E}{\mathfrak N} &=& C \frac{S\omega_0^2}{4J}(\gamma_x^2\gamma_y^2 + \gamma_x^2\gamma_z^2 + \gamma_y^2\gamma_z^2 ),\\
\label{c}
C &=& \frac{SJ}{\omega_0^2 \mathfrak N} \sum_{\bf q}\frac{\left( Q_{\bf q}^{xx} - Q_{\bf q}^{yy} \right)^2 - 4 \left( Q_{\bf q}^{xy} \right)^2 }{4\epsilon_{\bf q}},
\end{eqnarray}
where $\gamma_i$ are direction cosines of the magnetization and components of the dipolar tensor in Eq.~(\ref{c}) are taken relative to cubic axes. The constant $C$ can be calculated numerically using the technique of dipolar sums computation (see, e.g., Ref.~\cite{sums} and references therein). We obtain in accordance with Refs.~\cite{tes,kef_an} that it is positive for SCL ($C\approx0.012$) and negative for FCCL ($C\approx-0.005$) and BCCL ($C\approx-0.04$). Then, for SCL an edge of the cube is easy direction whereas a body diagonal of the cube is easy direction for FCCL and BCCL.

SCL is considered in detail in the next three subsections (\ref{bd}, \ref{ld} and \ref{re}). Calculations and results for FCCL and BCCL resemble those for SCL. They are discussed in Sec.~\ref{fbc}.

We study now separately bubble diagrams shown in Fig.~\ref{diag} a) and b), and the loop diagram presented in Fig.~\ref{diag} c). 

\subsection{Bubble diagrams}
\label{bd}

Let us begin with the diagram shown in Fig.~\ref{diag} a). It appears from three-magnons terms (\ref{h3}) and gives zero. To demonstrate this we make all possible couplings of two operators $a$ and $a^\dagger$ in Eq.~(\ref{h3}): 
\begin{eqnarray}
&&\sum_{\bf q}
\left[
\frac{E_{\bf q}(1+2N_{\bf q}) - \epsilon_{\bf q}}{\epsilon_{\bf q}}
\left\{a_{\bf 0}^\dagger (Q_{\bf q}^{xz}+iQ_{\bf q}^{yz}) 
+ 
a_{\bf 0} (Q_{\bf q}^{xz}-iQ_{\bf q}^{yz})\right\}\right.\nonumber\\
&&\left.
{}- 
a_{\bf 0}^\dagger \frac{B^*_{\bf q}(Q_{\bf q}^{xz}-iQ_{\bf q}^{yz})}{\epsilon_{\bf q}}
(1+2N_{\bf q})
-
a_{\bf 0}\frac{B_{\bf q}(Q_{\bf q}^{xz}+iQ_{\bf q}^{yz})}{\epsilon_{\bf q}}
(1+2N_{\bf q})
\right],
\label{h3mod}
\end{eqnarray}
where $N_{\bf q}=(e^{\epsilon_{\bf q}/T}-1)^{-1}$ is the Plank's function. Expression (\ref{h3mod}) is zero because $Q_{\bf q}^{xz} = -Q_{\bf q'}^{xz}$, $Q_{\bf q}^{yz} = -Q_{\bf q'}^{yz}$, $E_{\bf q} = E_{\bf q'}$ and $B_{\bf q} = B_{\bf q'}$, where ${\bf q}=(q_x,q_y,q_z)$ and ${\bf q'}=(-q_x,-q_y,q_z)$.

The Hartree-Fock diagram presented in Fig.~\ref{diag} b) comes from ${\cal H}_4$-terms given by Eq.~(\ref{h4}). After simple calculations we obtain for contribution to $\Omega(\omega,{\bf k})$ from the four-magnons terms:
\begin{eqnarray}
\label{o4}
\Omega^{(4)}(\omega,{\bf k}) &=& 
\frac{E_{\bf k}}{\mathfrak N} \sum_{\bf q}\left[\frac{E_{\bf q}(1+2N_{\bf q}) - \epsilon_{\bf q}}{\epsilon_{\bf q}} \left\{(J_{\bf k} - J_0 + J_{\bf q} - J_{\bf k-q}) 
+ Q_{\bf q}^{xx} - Q_{\bf k-q}^{zz} 
+ \frac{Q_{\bf k}^{xx} + Q_{\bf k}^{yy}}{2}
- Q_0^{zz}\right\} 
\right.
\nonumber\\
&&\left.
{}+ \frac{|B_{\bf q}|^2(1+2N_{\bf q})}{S\epsilon_{\bf q}} 
\right]
+ 
\frac{|B_{\bf k}|^2}{S \mathfrak N} \sum_{\bf q}\frac{E_{\bf q}(1+2N_{\bf q}) - \epsilon_{\bf q}}{\epsilon_{\bf q}}
- \frac{1}{\mathfrak N}\sum_{\bf q}J_{\bf k-q}\frac{{\rm Re}(B_{\bf q}^*B_{\bf k})(1+2N_{\bf q})}{\epsilon_{\bf q}}.
\end{eqnarray}
At zero temperature $N_{\bf q}=0$ in Eq.~(\ref{o4}). In this case the spectrum is renormalized by quantum fluctuations only. As the temperature increases, corrections from terms in Eq.~(\ref{o4}) containing $N_{\bf q}$ become larger. They exceed terms in Eq.~(\ref{o4}) not containing $N_{\bf q}$ above a certain temperature. We find below that this temperature is of the order of $(D\omega_0^2)^{1/3}$. Regimes of $T\ll (D\omega_0^2)^{1/3}$ and $T\gg (D\omega_0^2)^{1/3}$ are called below as quantum and thermal ones, respectively. It is convenient to discuss them separately.

\underline{\it $T\ll (D\omega_0^2)^{1/3}$}. To our knowledge corrections to the spin-wave spectrum in this regime have not been considered before. Taking $N_{\bf q}=0$ in Eq.~(\ref{o4}) one notes that summation over large momenta is essential and 
$(E_{\bf q}-\epsilon_{\bf q})/\epsilon_{\bf q} \approx |B_{\bf q}|^2/(2\epsilon_{\bf q}^2)$. 
As a result, we obtain in the leading order of $\omega_0$:
\begin{eqnarray}
\label{o4t<}
\Omega^{(4)}(\omega,{\bf k}) &=& 
-Dk^2 \frac{\omega_0^2}{J}\sin^4\theta_{\bf k}\cos^22\phi_{\bf k} X 
+ Dk^2 \frac{1}{\mathfrak N} \sum_{\bf q}\frac{|B_{\bf q}|^2}{S\epsilon_{\bf q}} 
+ \omega_0\sin^2\theta_{\bf k} \frac{1}{\mathfrak N} \sum_{\bf q}\frac{|B_{\bf q}|^2}{2\epsilon_{\bf q}},\\
\label{x}
X &=& \frac{SJ}{\omega_0 \mathfrak N}\sum_{\bf q}\frac{Q_{\bf q}^{xx} - Q_{\bf q}^{yy}}{4\epsilon_{\bf q}}\cos q_x \approx 0.013,
\end{eqnarray}
where $\phi_{\bf k}$ is the azimuthal angle of $\bf k$ and $X$ is calculated numerically. The first and the last two terms in Eq.~(\ref{o4t<}) stem, respectively, from the last and the first terms in Eq.~(\ref{o4}). Dependence of $\Omega^{(4)}(\omega,{\bf k})$ on $\bf k$ differ drastically from that of the bare spectrum (\ref{spec0}). The first and the second terms in Eq.~(\ref{o4t<}) give corrections to the spin-wave stiffness. It is seen that the first term depends on azimuthal angle $\phi_{\bf k}$. Such a dependence is allowed by the symmetry because the first term can be expressed via the quadratic invariant: $k^2\sin^4\theta_{\bf k}\cos^22\phi_{\bf k}=(k_x^2-k_y^2)^2/k^2$. Essentially, the second term does not depend on direction of $\bf k$. Thus, it plays the predominant part when $\sin\theta_{\bf k}=0$ and $k\ll\omega_0/D$. In this region $\Omega^{(4)}(\omega,{\bf k})$ is proportional to $k^2$, whereas $\epsilon_{\bf k}^2\propto k^4$, and the renormalized spectrum appears to be linear rather than quadratic. The last term remains finite as $k\to0$ and $\sin\theta_{\bf k} \ne 0$ and gives a contribution to the spin-wave gap. 

We stress that in all sums in Eq.~(\ref{o4t<}) summation over large $\bf q$ is essential and it is important to take into account orientation of quantized axis relative to cubic axes.

\underline{\it $T\gg(D\omega_0^2)^{1/3}$}. Calculation of the terms in Eq.~(\ref{o4}) containing $N_{\bf q}$ leads us to the following expression in the leading order of $\omega_0$:
\begin{eqnarray}
\label{o4t>}
\Omega^{(4)}(\omega,{\bf k}) &=& 
- \left(Dk^2\right)^2\frac2S W(T) 
- Dk^2\omega_0\sin^2\theta_{\bf k} [W(T)+V(T)]
\nonumber\\
&&{}
+ 
\left( Dk^2 + \frac{S\omega_0}{2}\sin^2\theta_{\bf k} \right)
\left( 
2\omega_0 {\cal N}_z V(T) 
+ 
\frac{1}{\mathfrak N} \sum_{\bf q}\frac{|B_{\bf q}|^2}{S\epsilon_{\bf q}}(1 + 2N_{\bf q})
\right),\\
\label{w}
W(T) &=& \frac{1}{\mathfrak N} \sum_{\bf q} \frac{J_{\bf 0} - J_{\bf q}}{J_{\bf 0}}  N_{\bf q} \approx w \left(\frac TD\right)^{5/2},\\
\label{v}
V(T) &=& \frac{1}{\mathfrak N} \sum_{\bf q} N_{\bf q} \approx v \left(\frac TD\right)^{3/2},
\end{eqnarray}
where $w = 0.01$ and $v = 0.06$. The first term in Eq.~(\ref{o4}) gives the largest thermal contribution to $\Omega^{(4)}(\omega,{\bf k})$. Let us make sure now that the thermal corrections given by Eq.~(\ref{o4t>}) are larger than the quantum ones (\ref{o4t<}) at $T\gg(D\omega_0^2)^{1/3}$. Indeed, the first two terms in Eq.~(\ref{o4t>}) are greater than the first term in Eq.~(\ref{o4t<}) at such temperatures. Let us compare the third term in Eq.~(\ref{o4t>}) with the second and the third terms in Eq.~(\ref{o4t<}). If ${\cal N}_z\ne0$ the first term in the second brackets in Eq.~(\ref{o4t>}) are larger than the second one and the last two terms in Eq.~(\ref{o4t<}) when $T\gg(D\omega_0^2)^{1/3}$. If ${\cal N}_z = 0$ only the second term remains in the second brackets in Eq.~(\ref{o4t>}). It is proportional to $\omega_0^2$ and $TD(\omega_0/D)^{3/2}$ at $T\ll\sqrt{\omega_0D}$ and $T\gg\sqrt{\omega_0D}$, respectively. We demonstrate below that there is a correction from the loop diagram that cancel the $T$-dependent contribution from this term. That is why we keep the unity in the factor $(1 + 2N_{\bf q})$ under the sum. As a result if ${\cal N}_z = 0$ the third term in Eq.~(\ref{o4t>}) is equal to the sum of the last two terms in Eq.~(\ref{o4t<}).

The first two terms in Eq.~(\ref{o4t>}) result in the following renormalization of constants $D$ and $\omega_0$ in Eq.~(\ref{spec0}):
\begin{equation}
\label{ren}
D \mapsto D\left[1 - \frac 1S W(T)\right],\qquad
\omega_0 \mapsto \omega_0 \left[1 - \frac 1S V(T)\right].
\end{equation}
At the same time the last term in Eq.~(\ref{o4t>}) has another structure. It contains the isotropic correction to the spin-wave stiffness. As it is explained above, such a correction plays the predominant part when $\sin\theta_{\bf k}=0$ and $k \ll \omega_0/D$. The last term in Eq.~(\ref{o4t>}) contributes also to the spin-wave gap.

Notably, thermal contributions to the isotropic correction to the spin-wave stiffness and to the gap depend on the demagnetizing factor ${\cal N}_z$. This surprising feature is attributed to the long-range nature of the dipolar interaction.

It should be noted that only the first two terms in Eq.~(\ref{o4t>}) have been obtained in Ref.~\cite{rahman}, where only the regime was studied in which the thermal fluctuations prevail. The origin of this discrepancy is in the following. First, it was discussed the special geometry of the sample in Ref.~\cite{rahman}: a macroscopic slab of finite thickness $d$ with square faces of length $L\to\infty$; the magnetization is parallel to the surface. Demagnetizing factor ${\cal N}_z$ is zero under these conditions. Second, to put it in our terminology, the anomalous Green's functions $F$ and $F^\dagger$ were discarded in all sums on the path leading to Eq.~(\ref{o4}). As a result the terms in Eq.~(\ref{o4}) containing $B_{\bf q}$ under sums were missed. In particular the last term under the first sum in Eq.~(\ref{o4}) was discarded that gives rise to the second term in the second brackets in Eq.~(\ref{o4t>}), i.e., that leads to the spin-wave gap and to the isotropic correction to the spin-wave stiffness if ${\cal N}_z=0$. That is why only renormalization of constants $D$ and $\omega_0$ given by Eqs.~(\ref{ren}) was obtained in Ref.~\cite{rahman} from ${\cal H}_4$-terms.

We point out that summation over small $\bf q$ is important in Eqs.~(\ref{w}) and (\ref{v}). Thus terms in Eq.~(\ref{o4t>}) containing $W(T)$ and $V(T)$ appears also in the case of BCCL and FCCL. The difference is in constants $w$ and $v$ only. 

It should be noted that there is also a renormalization of $D$ and $\omega_0$ similar to that given by Eq.~(\ref{ren}) in quantum regime. But corresponding terms in $\Omega^{(4)}(\omega,{\bf k})$ are much smaller than those presented in Eq.~(\ref{o4t<}) and we omitted them.

\subsection{Loop diagram}
\label{ld}

We turn now to the loop diagram shown in Fig.~\ref{diag} c). It appears from ${\cal H}_3$-terms (\ref{h3}) in the Hamiltonian. As a result of simple but tedious calculations some details of which are presented in Appendix~\ref{loop} we obtain for the contribution to $\Omega(\omega,{\bf k})$ from this diagram:
\begin{eqnarray}
\label{o3t}
\Omega^{(3)}(\omega,{\bf k}) &=& -Dk^2 \frac{\omega_0^2}{J} \sin^22\theta_{\bf k} U(\omega,{\bf k},T)
-
\left(Dk^2+\frac{S\omega_0}{2}\sin^2\theta_{\bf k}\right)
\nonumber\\
&&{} \times
\frac{S}{\mathfrak N} \sum_{\bf q} 
\frac{\left[ 6E_{\bf q}^2 + 2|B_{\bf q}|^2 + 2\epsilon_{\bf q}^2 \right] \left[\left( Q_{\bf q}^{xz} \right)^2 + \left( Q_{\bf q}^{yz} \right)^2\right] - 8E_{\bf q} {\rm Re}\left(B_{\bf q}[Q_{\bf q}^{xz} + iQ_{\bf q}^{yz}]^2\right)}{8\epsilon_{\bf q}^3} (1 + 2N_{\bf q}),\\
\label{u}
U(\omega,{\bf k},T) &=& \frac{SJ}{4 \mathfrak N}\sum_{\bf q}
\left[
\frac{{\rm Re}(B_{\bf q}B^*_{\bf q-k}) + E_{\bf q}E_{\bf q-k} - \epsilon_{\bf q}^2}{\epsilon_{\bf q}[(\epsilon_{\bf q}+\epsilon_{\bf q-k})^2 - (\omega+i\delta)^2]}(1 + 2N_{\bf q})
\right.
\nonumber\\
&&{}+\left.
\frac{2[{\rm Re}(B_{\bf q}B^*_{\bf q-k}) + E_{\bf q}E_{\bf q-k} + \epsilon_{\bf q}\epsilon_{\bf q-k}][\epsilon_{\bf q}-\epsilon_{\bf q-k}]}{[(\epsilon_{\bf q}+\epsilon_{\bf q-k})^2 - (\omega+i\delta)^2][(\epsilon_{\bf q}-\epsilon_{\bf q-k})^2 - (\omega+i\delta)^2]}\left(N_{\bf q-k} - N_{\bf q}\right)
\right],
\end{eqnarray}
where we set $k=0$ in the sum of the second term in Eq.~(\ref{o3t}). Dimensionless function $U(\omega,{\bf k},T)$ contains the infrared singularity obtained previously by Rahman and Mills. \cite{rahman} At $k,\omega\to 0$ small $\bf q$ give the main contribution to the sum in Eq.~(\ref{u}). As a result one leads to estimations 
\begin{eqnarray}
{\rm Re} [U(\omega,{\bf k},T)] &\sim& \frac{T}{\sqrt{D\omega_0}} + \sqrt{\frac{\omega_0}{D}} \ln (k^2 + [\omega/\omega_0]^2),\\
{\rm Im} [U(\omega,{\bf k},T)] &\sim& \sqrt{\frac{\omega_0}{D}} 
\left(
1 + \frac{T}{\omega_0\sqrt{k^2 + [\omega/\omega_0]^2}}
\right).
\end{eqnarray}

\subsection{Resultant expressions}
\label{re}

Using Eqs.~(\ref{o4t<}), (\ref{o4t>}) and (\ref{o3t}) one can derive the resultant expression for $\Omega(\omega,{\bf k})$. Quantum and thermal regimes should be considered: 
$T \ll (D\omega_0^2)^{1/3}$ and
$T \gg (D\omega_0^2)^{1/3}$.

\underline{\it $T \ll(D\omega_0^2)^{1/3}$}. From Eqs.~(\ref{o4t<}) and (\ref{o3t}) we obtain
\begin{eqnarray}
\label{ot1}
\Omega(\omega,{\bf k}) &=& -Dk^2 \frac{\omega_0^2}{J} \left(X  \sin^4\theta_{\bf k}\cos^22\phi_{\bf k} 
- Y \right)
+ Y \frac{S\omega_0^3}{2J}\sin^2\theta_{\bf k},\\
\label{y}
Y &=& \frac{SJ}{\omega_0^2 \mathfrak N} \sum_{\bf q}\frac{\left| Q_{\bf q}^{xx} - 2iQ_{\bf q}^{xy} - Q_{\bf q}^{yy} \right|^2 - 4\left[ \left( Q_{\bf q}^{xz} \right)^2 + \left( Q_{\bf q}^{yz} \right)^2\right]}{4\epsilon_{\bf q}} \approx 0.012,
\end{eqnarray}
where $X$ is given by Eq.~(\ref{x}), the first term in the brackets in Eq.~(\ref{ot1}) originates from $\Omega^{(4)}(\omega,{\bf k})$ and to the rest terms contribute both $\Omega^{(3)}(\omega,{\bf k})$ and $\Omega^{(4)}(\omega,{\bf k})$. The gap which square is given by the last term in Eq.~(\ref{ot1}) is proportional to $\omega_0^{3/2}\sin\theta_{\bf k}$. The logarithmic divergence of $U(\omega,{\bf k},T)$ is screened by the gap and contribution from the first term in Eq.~(\ref{o3t}) was omitted in Eq.~(\ref{ot1}) being of the form $k^2\sin^22\theta_{\bf k}[\omega_0^2\sqrt{\omega_0/D}\ln(\omega_0/D) + T\omega_0^{3/2}]$.

It is easy to see that $Q_{\bf q}^{xz}$ and $Q_{\bf q}^{yz}$ can be replaced by $Q_{\bf q}^{xy}$ in Eq.~(\ref{y}). After this transformation $Y$ appears to be equal to $C$ given by Eq.~(\ref{c}). This coincidence is not accidental. As is discussed in Introduction, it looks reasonable that dipolar anisotropy given by Eq.~(\ref{an}) should be accompanied with a spin-wave gap. The relation between the dipolar anisotropy and terms in Eq.~(\ref{ot1}) containing $Y$ can be shown in the following non-rigorous way. Let us discuss large spins and try to take into account the dipolar anisotropy (\ref{an}) phenomenologically by adding to the microscopic Hamiltonian (\ref{ham}) the following expression [cf. Eq.~(\ref{an})]:
\begin{equation}
\label{anphen}
C\frac{\omega_0^2}{4S^3J}\sum_l\left[\left(S_l^x\right)^2\left(S_l^y\right)^2 + \left(S_l^x\right)^2\left(S_l^z\right)^2 + \left(S_l^y\right)^2\left(S_l^z\right)^2\right].
\end{equation}
This term after Dyson-Maleev transformation (\ref{md}) gives the contribution $C\omega_0^2/(2J)a^\dagger_{\bf k}a_{\bf k}$ to the bilinear part (\ref{h2}) of the Hamiltonian that in turn leads to the shift $E_{\bf k}\mapsto E_{\bf k} + C\omega_0^2/(2J)$. Using this renormalization of $E_{\bf k}$, equality $C=Y$ and Eq.~(\ref{spec1}) for the spectrum one recovers corrections in Eq.~(\ref{ot1}) containing $Y$. This consideration does not work if $S\le3/2$ because Eq.~(\ref{anphen}) is a constant in this case.

\underline{\it $T \gg (D\omega_0^2)^{1/3}$}. One has from Eqs.~(\ref{o4t>}) and (\ref{o3t}):
\begin{eqnarray}
\label{ot2}
\Omega(\omega,{\bf k}) &=& 
-\left(Dk^2\right)^2\frac2S W(T) 
- 
Dk^2\omega_0 \left( \sin^2\theta_{\bf k}[W(T) + V(T)]
-
2{\cal N}_z V(T) 
-
Y\frac{\omega_0}{J}
\right) \nonumber\\
&&{}
+
\frac{S\omega_0^2}{2}\sin^2\theta_{\bf k} \left( 
2 {\cal N}_z V(T) + Y\frac{\omega_0}{J}\right),
\end{eqnarray}
where $W(T)$, $V(T)$ and $Y$ are given by Eqs.~(\ref{w}), (\ref{v}) and (\ref{y}), respectively. It is seen from Eq.~(\ref{ot2}) that the gap is proportional to $\omega_0 T^{3/4}\sin\theta_{\bf k}$ and $\omega_0^{3/2}\sin\theta_{\bf k}$ if ${\cal N}_z \ne 0$ and ${\cal N}_z = 0$, respectively. It screens the divergence of $U(\omega,{\bf k},T)$. When ${\cal N}_z = 0$ the spin-wave gap has the same form as in the quantum regime. 

It should be noted that the second term in the second brackets of Eq.~(\ref{o4t>}) and the sum in the second term of Eq.~(\ref{o3t}) are proportional to $\omega_0^2$ and $TD(\omega_0/D)^{3/2}$ at $T\ll\sqrt{\omega_0D}$ and $T\gg\sqrt{\omega_0D}$, respectively. At small $T$ they result in terms proportional to the constant $Y$ in Eq.~(\ref{ot1}). The temperature corrections from these terms which were large at $T\gg\sqrt{\omega_0D}$ cancel each other and the terms in Eq.~(\ref{ot1}) proportional to $Y$ survive in the thermal regime.

\subsection{Face- and body-centered cubic lattices}
\label{fbc}

In the case of FCCL and BCCL arguments similar to those presented at the beginning of Sec.~\ref{bd} leads us to the conclusion that the diagram shown in Fig.~\ref{diag} a) is zero. The general expressions (\ref{o4}) and (\ref{o3t}) for $\Omega^{(4)}(\omega,{\bf k})$ and $\Omega^{(3)}(\omega,{\bf k})$ are valid for FCCL and BCCL. The results for $\Omega(\omega,{\bf k})$ differ slightly from Eqs.~(\ref{ot1}) and (\ref{ot2}). At $T \ll (D\omega_0^2)^{1/3}$ one obtains
\begin{equation}
\label{otfb}
\Omega(\omega,{\bf k}) = -Dk^2 \frac{\omega_0^2}{J} \left( X_1  \sin^4\theta_{\bf k} + X_2 \sin^3\theta_{\bf k} \cos\theta_{\bf k}\sin3\phi_{\bf k} 
- Y \right)
+ Y \frac{S\omega_0^3}{2J}\sin^2\theta_{\bf k},
\end{equation}
where we assume that $z$- and $x$- axes are directed along $(1,-1,1)$ and $(1,1,0)$, respectively, for BCCL one has
\begin{equation}
\label{xb}
X_2 = \sqrt2X_1 = \frac{\sqrt2}{3} \frac{SJ}{\omega_0 \mathfrak N}\sum_{\bf q}\frac{Q_{\bf q}^{xy}}{\epsilon_{\bf q}}\cos(q_x + q_y + q_z)
\approx 0.017, 
\end{equation}
for FCCL we obtain
\begin{eqnarray}
\label{x1f}
X_1 &=& \frac{SJ}{\omega_0 \mathfrak N}\sum_{\bf q}\frac{ (4Q_{\bf q}^{xy} + Q_{\bf q}^{xx} ) \cos(k_x+k_y) - Q_{\bf q}^{xx} \cos(k_y+k_z) }{24\epsilon_{\bf q}} 
\approx 0.003, \\
\label{x2f}
X_2 &=& \frac{SJ}{\omega_0 \mathfrak N}\sum_{\bf q}\frac{ (2Q_{\bf q}^{xy} - Q_{\bf q}^{xx} ) \cos(k_x+k_y) + Q_{\bf q}^{xx} \cos(k_y+k_z) }{6\sqrt2\epsilon_{\bf q}} 
\approx 0.002,
\end{eqnarray}
and constant $Y$ is given by Eq.~(\ref{y}) where components of the dipolar tensor are taken now within the coordinate frame with $z$-axis directed along the body diagonal of the cube. Numerical calculation gives
\begin{equation}
\label{yfb}
Y\approx \left\{
\begin{array}{ll}
0.03 & \mbox{  for BCCL,}\\
0.003 & \mbox{  for FCCL.}
\end{array}
\right.
\end{equation}
We see comparing Eqs.~(\ref{ot1}) and (\ref{otfb}) that the term proportional to the square invariant in the case of SCL is replaced by the term which can be expressed via triangular invariant ${\rm Im}([k_x + ik_y]^3)/k$ for BCCL and FCCL. This result is quite natural because the quantized axis in BCCL and FCCL, body diagonal of the cube, is triad axis.

In thermal regime Eq.~(\ref{ot2}) is valid for FCCL and BCCL, where $W(T)$ and $V(T)$ are given by Eqs.~(\ref{w}) and (\ref{v}), respectively. The numerical constants $w$ and $v$ in these expressions are following: $w=0.09$ and 0.03 for BCCL and FCCL, respectively; $v=0.23$ and 0.12 for BCCL and FCCL, respectively.

The relation between dipolar anisotropy (\ref{an}) and terms containing $Y$ in Eq.~(\ref{otfb}) can be demonstrated (non-rigorously) as it is done above for SCL. After addition expression (\ref{anphen}) to Hamiltonian (\ref{ham}) its bilinear part (\ref{h2}) acquires new term $-C\omega_0^2/(3J)a^\dagger_{\bf k}a_{\bf k}$ that in turn leads to the following renormalization: $E_{\bf k}\mapsto E_{\bf k} - C\omega_0^2/(3J)$. Considering this renormalization, using Eq.~(\ref{spec1}) for the spectrum and taking into account the equality
\begin{equation}
C = -\frac32 Y
\end{equation}
that follows from Eqs.~(\ref{c}) and (\ref{y}) one recovers terms containing $Y$ in Eq.~(\ref{otfb}).

\section{Discussions}
\label{disc}

As is seen from Eqs.~(\ref{ot1}) and (\ref{ot2}), the spin-wave gap screens the logarithmic divergence of the first term in Eq.~(\ref{o3t}) making it negligibly small. Let us discuss briefly further $1/S$-contributions. Bearing in mind that imaginary part of $U(\omega,{\bf k},T)$ given by Eq.~(\ref{u}) possesses power-law singularity $i\sqrt{\omega_0/D}T/\omega$ one could expect that divergences of further $1/S$-contributions to the stiffness contain powers of $\sqrt{\omega_0/D}T/\omega$. The spin-wave gap screens such singularities and the corresponding terms turn into powers of $T/\omega_0$. Thus, it is clear that at $T\ll\omega_0$ the analysis of higher order $1/S$-terms is not required. But such an analysis is necessary at $T\gg\omega_0$ that is out of the scope of the present paper.

Let us turn to the problem of infrared divergence of longitudinal spin fluctuations obtained in Ref.~\cite{top}. The longitudinal spin Green's function is given by 
$
\chi_\|(\omega,{\bf k}) = i\int_0^\infty dt e^{i\omega t}\langle [S^z_{\bf k}(t), S^z_{\bf k}(0)] \rangle.
$
Within the first order of $1/S$ one has $\chi_\|(\omega,{\bf k}) = 4U(\omega,{\bf k},T)/SJ$. It is clear from the above consideration that the spin-wave gap screens the divergence of $U(\omega,{\bf k},T)$. The analysis of further $1/S$-terms is not required at $T\ll\omega_0$. However such an analysis is necessary at $T\gg\omega_0$. 

Finally, we discuss the possibility to verify by inelastic neutron scattering theoretical results obtained above. Unfortunately to the best of our knowledge there is no suitable substance for the corresponding nowaday experiment. The most investigated compounds in which dipolar interaction plays important role ($\omega_0\sim D$) are europium monochalcogenides EuO and EuS. But according to the analysis of Ref.~\cite{kasuya}, one-ion anisotropy is very large in these materials and seems to screen all the peculiarities discussed above. Seemingly, there is no such problem in chalcogenide spinel CdCr$_2$Se$_4$. According to estimations made in Ref.~\cite{luz} the anisotropy field observed in this substance is of the order of dipolar anisotropy. At the same time $\omega_0/D\approx 10^{-3}$ in CdCr$_2$Se$_4$ and extremely small impulses transfer of the order of $10^{-3}\mbox{ \AA}^{-1}$ should be achieved in the corresponding neutron experiment. It is hardly possible nowadays.

\section{Conclusion}
\label{con}
In the present paper an evaluation is carried out of the real part of first $1/S$-corrections to the spin-wave spectrum of a cubic ferromagnet with dipolar forces. In accordance with previous results \cite{rahman} we obtain thermal renormalization (\ref{ren}) of constants $D$ and $\omega_0$ in the bare spectrum (\ref{spec0}). Besides, a number of previously unknown features are revealed. We observe terms which depend on azimuthal angle of the momentum $\bf k$. They are proportional to the square invariant $|k_x^2 - k_y^2|/k$ in the case of simple cubic lattice (SCL) and to the triangular invariant ${\rm Im}([k_x + ik_y]^3)/k^2$ in face-centered cubic lattice (FCCL) and body-centered cubic lattice (BCCL), where we assume that $z$-axis is directed along magnetization. This difference is accounted for the fact that dipolar interaction gives rise to the anisotropy: edges of a cube are easy directions in SCL and body diagonals of a cube are easy directions in BCCL and FCCL. \cite{tes,kef_an} It is shown that there is a correction to the spin-wave spectrum proportional to $k$ which does not depend on direction of $\bf k$ (isotropic correction). Thus the spectrum appears to be linear in $k$ rather than quadratic [cf. Eq.~(\ref{spec0})] when $k\ll\omega_0/D$ and $\sin\theta_{\bf k}=0$, where $\theta_{\bf k}$ is the polar angle of $\bf k$. In particular, we observe a spin-wave gap proportional to $\sin\theta_{\bf k}$. We demonstrate a relation between the dipolar anisotropy and the gap and isotropic term in the spectrum.

Notably, thermal contribution from the Hartree-Fock diagram shown in Fig.~\ref{diag} b) to the isotropic correction as well as to the spin-wave gap are proportional to the demagnetizing factor in the direction of the domain magnetization. Then the spin-wave excitations with wavelengths much smaller than the size of the domain appear to "feel" the domain shape. This nontrivial behavior is attributed to the long-range nature of the dipolar interaction.

It is demonstrated that the gap screens singularities of perturbation corrections to the spin-wave stiffness \cite{rahman} and longitudinal dynamical spin susceptibility \cite{top} obtained before. We show that higher order $1/S$-corrections to these quantities are small at $T\ll\omega_0$, where $\omega_0$ is the characteristic dipolar energy given by Eq.~(\ref{o0}). However the analysis of entire perturbation series is still required to derive the spectrum and the susceptibility when $T\gg\omega_0$. Such an analysis is out of the scope of the present paper. One could expect large renormalization of the longitudinal dynamical susceptibility at $T\gg\omega_0$ as the nontrivial power-law dependence of $\chi_\|(\omega\to0) \sim (i/\omega)^{0.28}$ was obtained experimentally at such temperatures. \cite{luz}

\begin{acknowledgments}

I am grateful to A.V.\ Lazuta, S.V.\ Maleyev and A.G.\ Yashenkin for numerous discussions. This work was supported by Russian Science Support Foundation, grant of President of Russian Federation MK-4160.2006.2, RFBR (Grant Nos. 06-02-16702 and 00-15-96814) and Russian Programs "Quantum Macrophysics", "Strongly correlated electrons in semiconductors, metals, superconductors and magnetic materials" and "Neutron Research of Solids".

\end{acknowledgments}

\appendix

\section{Calculation of $\Omega^{(3)}(\omega,{\bf k})$}
\label{loop}

\begin{figure}
\centering
\includegraphics[scale=0.4]{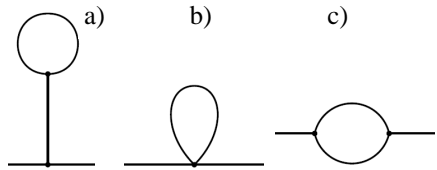}
\caption{Diagrams of the first order of $1/S$. Diagrams a) and c) stem from three-magnons terms (\ref{h3}) in the Hamiltonian whereas b) comes from four-magnons terms (\ref{h4}).
\label{diag}} 
\end{figure}

We present in this appendix some details of calculation of $\Omega^{(3)}(\omega,{\bf k})$ that is a contribution to $\Omega(\omega,{\bf k})$ given by Eq.~(\ref{o}) from the loop diagram shown in Fig.~\ref{diag} c). This diagram appears from ${\cal H}_3$-terms (\ref{h3}) in the Hamiltonian. As a result of simple calculations we lead to quit a cumbersome expression:
\begin{subequations}
\label{o3}
\begin{eqnarray}
\Omega^{(3)}(\omega,{\bf k}) &=& 
-Dk^2 \omega_0^2 \sin^22\theta_{\bf k} \frac{S}{4 \mathfrak N} \sum_{{\bf q}_1,{\bf q}_2} \langle a_{\bf 1} a^\dagger_{\bf 1-k}, a_{-\bf 2}a^\dagger_{\bf k-2} \rangle_\omega\\
&&{}
-\left(Dk^2+\frac{S\omega_0}{2}\sin^2\theta_{\bf k}\right)\frac{S}{2 \mathfrak N}
\sum_{{\bf q}_1,{\bf q}_2}\left[ \langle M_{\bf 1}a^\dagger_{\bf 1-k}, M_{-\bf 2} a_{\bf 2-k}\rangle_\omega
+
\langle M_{\bf 1} a_{\bf k-1}, M_{-\bf 2}a^\dagger_{\bf k-2} \rangle_\omega \right]\\
&&{}
+\omega\frac{S}{2 \mathfrak N}
\sum_{{\bf q}_1,{\bf q}_2}\left[ \langle M_{\bf 1}a^\dagger_{\bf 1-k}, M_{-\bf 2} a_{\bf 2-k}\rangle_\omega
-
\langle M_{\bf 1} a_{\bf k-1}, M_{-\bf 2}a^\dagger_{\bf k-2} \rangle_\omega \right]\\
&&{} 
- \omega\omega_0\sin2\theta_{\bf k}e^{-i\phi_{\bf k}} \frac{S}{4 \mathfrak N} 
\sum_{{\bf q}_1,{\bf q}_2} \left[ \langle a_{\bf 1}a^\dagger_{\bf 1-k}, M_{-\bf 2} a_{\bf 2-k}\rangle_\omega
-
\langle M_{\bf 1} a_{\bf k-1}, a_{-\bf 2}a^\dagger_{\bf k-2} \rangle_\omega \right]\\
&&{} 
- \omega\omega_0 \sin2\theta_{\bf k}e^{i\phi_{\bf k}} \frac{S}{4 \mathfrak N} 
\sum_{{\bf q}_1,{\bf q}_2} \left[ \langle M_{\bf 1} a^\dagger_{\bf 1-k}, a_{-\bf 2}a^\dagger_{\bf k-2} \rangle_\omega 
- 
\langle a_{\bf 1}a^\dagger_{\bf 1-k}, M_{-\bf 2} a^\dagger_{\bf k-2}\rangle_\omega \right]\\
&&{}
+ Dk^2 \omega_0 \sin2\theta_{\bf k}e^{i\phi_{\bf k}} \frac{S}{4 \mathfrak N}
\sum_{{\bf q}_1,{\bf q}_2} \left[ \langle M_{\bf 1} a^\dagger_{\bf 1-k}, a_{-\bf 2}a^\dagger_{\bf k-2} \rangle_\omega 
+ \langle a_{\bf 1}a^\dagger_{\bf 1-k}, M_{-\bf 2} a^\dagger_{\bf k-2}\rangle_\omega \right]\\
&&{}
+ Dk^2 \omega_0 \sin2\theta_{\bf k}e^{-i\phi_{\bf k}} \frac{S}{4 \mathfrak N}
\sum_{{\bf q}_1,{\bf q}_2} \left[ \langle a_{\bf 1}a^\dagger_{\bf 1-k}, M_{-\bf 2} a_{\bf 2-k}\rangle_\omega
+
\langle M_{\bf 1} a_{\bf k-1}, a_{-\bf 2}a^\dagger_{\bf k-2} \rangle_\omega \right]\\
&&{}
+ B_{\bf k}\frac{S}{2 \mathfrak N} \sum_{{\bf q}_1,{\bf q}_2} \langle M_{\bf 1}a_{\bf k-1}, M_{-\bf 2} a_{\bf 2-k}\rangle_\omega
+
B_{\bf k}^*\frac{S}{2 \mathfrak N} \sum_{{\bf q}_1,{\bf q}_2} \langle M_{\bf 1} a^\dagger_{\bf 1-k}, M_{-\bf 2}a^\dagger_{\bf k-2} \rangle_\omega,
\end{eqnarray}
\end{subequations}
where $\langle\dots\rangle_\omega$ denotes retarded Green's function, indexes $\bf q$ are dropped in expressions under  sums and combination
\begin{equation}
\label{m}
M_{\bf q} = a_{\bf q}(Q_{\bf q}^{xz} - iQ_{\bf q}^{yz}) + a^\dagger_{-\bf q}(Q_{\bf q}^{xz} + iQ_{\bf q}^{yz})
\end{equation}
is introduced. For the first glance expression (\ref{o3}) looks awesome. Meanwhile it becomes simpler if one takes advantage of the properties of the combination $M_{\bf q}$:
\begin{eqnarray}
\langle M_{\bf q}, a_{-\bf q}\rangle_\omega &=& -B_{\bf q}^*(Q_{\bf q}^{xz} - iQ_{\bf q}^{yz}) + (E_{\bf q} - \omega)(Q_{\bf q}^{xz} + iQ_{\bf q}^{yz}) \stackrel{q\ll1}{\approx} - \frac{\omega_0}{2}\frac{(Dq^2 - \omega) \sin2\theta_{\bf q}}{(\omega+i\delta)^2-\epsilon_{\bf q}^2}e^{i\phi_{\bf q}},\nonumber\\
\label{mprop}
\langle M_{\bf q}, a^\dagger_{\bf q}\rangle_\omega &=& 
-B_{\bf q}(Q_{\bf q}^{xz} + iQ_{\bf q}^{yz}) 
+ (E_{\bf q} + \omega)(Q_{\bf q}^{xz} - iQ_{\bf q}^{yz}) \stackrel{q\ll1}{\approx} - \frac{\omega_0}{2} \frac{(Dq^2 + \omega) \sin2\theta_{\bf q}}{(\omega+i\delta)^2-\epsilon_{\bf q}^2}e^{-i\phi_{\bf q}},\\
\langle M_{\bf q}, M_{-\bf q}\rangle_\omega &=& 
2E_{\bf q} \left[ \left(Q_{\bf q}^{xz}\right)^2 + \left(Q_{\bf q}^{yz}\right)^2 \right] 
- B_{\bf q}\left(Q_{\bf q}^{xz} + iQ_{\bf q}^{yz}\right)^2 
- B^*_{\bf q} \left( Q_{\bf q}^{xz} - iQ_{\bf q}^{yz} \right)^2 \stackrel{q\ll1}{\approx} \frac{\omega_0^2}{2}\frac{Dq^2 \sin^22\theta_{\bf q}}{(\omega+i\delta)^2-\epsilon_{\bf q}^2},\nonumber
\end{eqnarray}
where we use expressions (\ref{qsmall}) and (\ref{q0}) for components of the dipolar tensor. One can demonstrate that terms (c)--(h) in Eq.~(\ref{o3}) are small compared to $\Omega^{(4)}(\omega,{\bf k})$ given by Eqs.~(\ref{o4t<}) and (\ref{o4t>}). It is easy to see that summation over small momentum give the main contribution to these terms. Their evaluation using Eqs.~(\ref{mprop}) is a tedious but straightforward work. Terms (h) vanish as $k\to0$ or $\theta_{\bf k}\to 0,\pi$. As a result they are proportional to $k^2\omega_0^{5/2}\sin^4\theta_{\bf k}$ and $\omega_0^4\sin^4\theta_{\bf k}/k$ at $k\ll\sqrt{\omega_0/D}$ and $k\gg\sqrt{\omega_0/D}$, respectively. Sums in terms (f) and (g) vanish as $k\to0$ or $\theta_{\bf k}\to 0,\pi/2,\pi$. In the issue (f) and (g) appear to be small compared to $\Omega^{(4)}(\omega,{\bf k})$ being proportional to $k^4\omega_0^{3/2}\sin^2 2\theta_{\bf k}$ and $k^3\omega_0^2\sin^2 2\theta_{\bf k}$ at $k\ll\sqrt{\omega_0/D}$ and $k\gg\sqrt{\omega_0/D}$, respectively. Sums in (d) and (e) are proportional to $\omega$ and vanish as $\theta_{\bf k}\to 0,\pi/2,\pi$. Then, (d) and (e) can be also discarded being proportional to $\omega^2k\omega_0\sin^2 2\theta_{\bf k}$ and $\omega^2\omega_0^2\sin^2 2\theta_{\bf k}/k$ at $k\ll\sqrt{\omega_0/D}$ and $k\gg\sqrt{\omega_0/D}$, respectively. Term (c) is negligible being proportional to $\omega^2\omega_0^{3/2}$ and $\omega^2\omega_0^2/k$ at $k\ll\sqrt{\omega_0/D}$ and $k\gg\sqrt{\omega_0/D}$, respectively. The above evaluations are valid for $T=0$. It can be demonstrated that temperature corrections to (c)--(h) are also negligible compared to $\Omega^{(4)}(\omega,{\bf k})$.

As a result only terms (a) and (b) in Eq.~(\ref{o3}) should be taken into account. They contribute to the first and the second terms in Eq.~(\ref{o3t}), respectively. This result is valid for all types of cubic lattices because summation over small momenta give the main contribution to (c)--(h). We point out that only normal self-energy parts ($\Sigma$ and $\overline{\Sigma}$) contribute to terms (a) and (b).

Notice that term (a) is the only source of singularity in Eq.~(\ref{o3}). As is seen from Eqs.~(\ref{mprop}), the combination $M_{\bf q}$ is "soft": numerators of all the products containing $M_{\bf q}$ vanishes as $q,\omega\to0$.

\bibliography{dipfm}

\end{document}